\documentclass[iop,apjl,numberappendix,appendixfloats]{emulateapj}
\usepackage{graphicx}
\newcommand{\lya}{Ly$\alpha$}

\newcommand{\zzz}{$z\sim$ 3}

\newcommand{\vvgal}{VVDS 910298177}
\newcommand{\hi}{H\textsc{i}}

\begin{document}
\title{A New Constraint on the Physical Nature of Damped Lyman Alpha
Systems}

\author{J. Cooke\altaffilmark{1} \& J.M. O'Meara\altaffilmark{2}}

\altaffiltext{1}{Centre for Astrophysics and Supercomputing, Swinburne
University of Technology, Hawthorn, Vic 3122, Australia}
\altaffiltext{2}{Department of Physics, Saint Michael's College, One
Winooski Park, Colchester, VT 05439, USA}

\begin{abstract}

The formation and evolution of galaxies require large reservoirs of
cold, neutral gas.  The damped Lyman-$\alpha$ systems (DLAs), seen in
absorption toward distant quasars and gamma-ray bursts, are predicted
to be the dominant reservoirs for this gas.  Detailed properties of
DLAs have been studied extensively for decades with great success.
However, their size, fundamental in understanding their nature, has
remained elusive, as quasar and gamma-ray-burst sightlines only probe
comparatively tiny areas of the foreground DLAs.  Here, we introduce a
new approach to measure the full extent of DLAs in the sightlines
toward {\it extended} background sources.  We present the discovery of
a high-column-density (log $N$(\hi) = 21.1 $\pm0.4$ cm$^{-2}$) DLA at
$z\sim2.4$ covering 90--100\% of the luminous extent of a
line-of-sight background galaxy.  Estimates of the size of the
background galaxy range from a minimum of a few kpc$^2$, to $\sim$100
kpc$^2$, and demonstrate that high-column density neutral gas can span
continuous areas 10$^8$--10$^{10}$ times larger than previously
explored in quasar or gamma-ray burst sightlines.  The DLA presented
here is the first from a sample of DLAs in our pilot survey that
searches Lyman break and Lyman continuum galaxies at high
redshift. The low luminosities, large sizes, and mass contents
($\gtrsim$10$^6$--10$^{9}$ $M_\odot$) implied by this DLA and the
early data suggest that DLAs contain the necessary fuel for galaxies,
with many systems consistent with relatively massive, low-luminosity
primeval galaxies.

\end{abstract}

\keywords{galaxies: evolution --- galaxies: high redshift ---
intergalactic medium --- quasars: absorption lines}

\section{Introduction}\label{intro}

\noindent Damped \lya\ absorption systems (DLAs) contain the bulk of
the neutral hydrogen (\hi) in the Universe and play a dominant role in
cosmic star formation \citep{wolfe05,wolfire03}.  High-resolution
spectra of DLAs in bright quasar and gamma-ray burst sightines yield a
wealth of information, such as their chemical composition, ionization
states, and gas kinematics \citep[e.g.,][]{prochaska07}.  However,
these background sources probe areas of $<$0.01 parsec$^2$ at the DLA
redshift and, as a result, the size of DLAs has remained elusive for
$\sim$40 years.  Only by their spatial correlation with a known galaxy
population has the average mass of DLAs been constrained
\citep{gawiser01,adelberger03,bouche04,cooke06}.  The situation
improves dramatically when the background light source has a large
spatial extent, such as a galaxy.  Extended background sources can
distinguish between DLAs that have large or small spatial extents via
differences in their observed absorption-line depths and profiles,
reflecting their covering fractions and column densities,
respectively.  In this Letter, we present the detection of a
high-redshift DLA as the first result of a new program to determine
DLA spatial extents as a population by searching comparatively high
signal-to-noise ratio (S/N), low-resolution $z\gtrsim$ 2 galaxy
spectra.  We assume an $H=$ 70, $\Omega_M=$ 0.3, $\Omega_{\Lambda}=$
0.7 cosmology throughout this paper, and all magnitudes are in the AB
system \citep{fukugita96}.

\section{Data}

Our program utilizes the publicly available VLT Vimos Deep Survey
(VVDS) UltraDeep \footnote{cesam.lam.fr/vvds/vvds\_download.php}
\citep{lefevre03,lefevre13} using the 8.2m Very Large Telescope.  To
date, we searched for DLAs in 54 $z\sim$ 2--4 galaxy spectra that meet
Lyman continuum galaxy \citep[LCG;][]{cooke14} criteria.  The
VVDS-UltraDeep consists of deep, $\sim$18 hr spectroscopic
integrations ($\sim$1200 s per exposure) that result in S/N
$\sim$5--20 in the \lya\ forest region of the galaxy spectra to search
for absorption-line systems.  VVDS-UltraDeep provides a bluer
wavelength coverage ($\sim$3650--9350 \AA) as compared to the parent
VVDS survey necessary for $z\sim$ 2--4 DLA searches.  In addition, we
searched for DLAs in 260 $z\sim$ 2--4 Lyman break galaxy
\citep[LBG;][]{steidel95} spectra using LRIS \citep{oke95,steidel04}
and DEIMOS \citep{faber03} on the 10m Keck telescopes between 2001
March to 2014 June as primary or secondary science for several
programs (hereafter the Keck programs).  Details of the observations
and the color selection criteria are outlined in previous work
\citep{cooke05,cooke13}.

The Keck programs color select $R_{AB}\le$ 25.5 (or $i_{AB}\le$ 25.5)
galaxies following the standard Lyman break technique that assumes
little to no escaping Lyman continuum ($<$912\AA) flux.  Galaxies that
produce Lyman continuum flux have observed colors inconsistent with
Lyman break expectations with respect to their redshifts, but are
found within, and outside of, the standard color selection criteria
\citep{cooke14}.  As a result, the Keck programs are only sensitive to
galaxies with foreground DLAs having a combined color that remains
within the standard LBG selection criteria.  In contrast,
VVDS-UltraDeep selects $i_{AB}>$ 24.75 galaxies solely based on their
magnitude (i.e., no color selection).  We focus on the VVDS-UltraDeep
spectra that exhibit excess $u$-band flux (LCGs), with most galaxies
having colors outside of the standard LBG color selection criteria.
These spectra are more likely to include lower-redshift line-of-sight
systems with detectable emission as compared to LBGs.

We use the five-year stacked images (m$_{lim}\sim$ 27) of the
Canada-France-Hawaii Telescope Legacy Survey (CFHTLS) Deep Fields
high-quality $u^*$$g^\prime$$r^\prime$$i^\prime$
images\footnote{www.cfht.hawaii.edu/Science/CFHLS/} (seeing FWHM $<
0.75^{\prime\prime}$) for photometric and morphological analysis.  The
imaging and spectroscopic data were reduced using standard IRAF and
IDL procedures and the reduction pipelines of the facilities providing
the data.

\section{\vvgal}

\begin{figure}
\begin{center}
\includegraphics[width=1.0\textwidth]{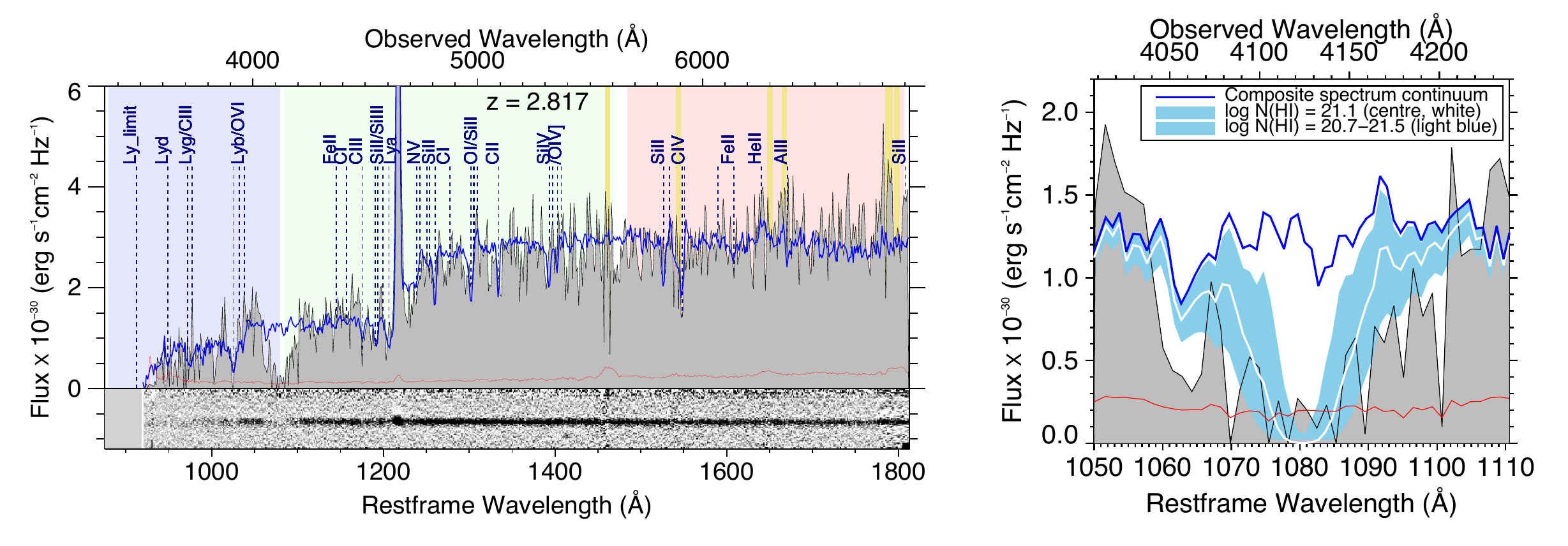}

\caption{Spectra of the $z=$ 2.818 galaxy \vvgal.  Left: the 1D
spectrum (black) and 1$\sigma$ error (red) are shown above zero flux,
with the 2D spectrum inserted below zero flux for comparison.  A
composite \zzz\ galaxy spectrum is overlaid (blue) with common atomic
transitions labeled (dashed vertical lines) and the positions of
bright night sky emission lines marked (thick yellow vertical lines).
The DLA is seen near 1080\AA, restframe (4120 \AA, observer-frame) and
the $u^*$$g^\prime$$r^\prime$ CFHT Megacam filter bandpasses are
indicated by the blue, green, and red shaded regions, respectively.
Right: zoom-in of the spectrum centered on the DLA.  A Voigt profile
fit to the absorption (white curve) yields log $N$(H\textsc{i}) = 21.1
$\pm$0.4 atoms cm$^{-2}$ (light blue shaded region).}

\label{f1}
\end{center}
\end{figure}

Figure~\ref{f1} presents the 1D and 2D spectra of the $z\sim$ 2.8
galaxy \vvgal\ from the VVDS-UltraDeep survey.  The spectral profile
is typical of $z\sim3$ LBGs with \lya\ in emission.  As seen in
Figure~\ref{f1}, the spectra show a strong absorption feature near
$\lambda \sim$ 4124\AA.  We interpret the feature as \hi\ \lya\
absorption.  Visual inspection of the broad width of the absorption
feature, which we refer to as EG1 for convenience, places it in the
DLA regime.

\begin{figure}
\begin{center}
\includegraphics[width=1.0\textwidth]{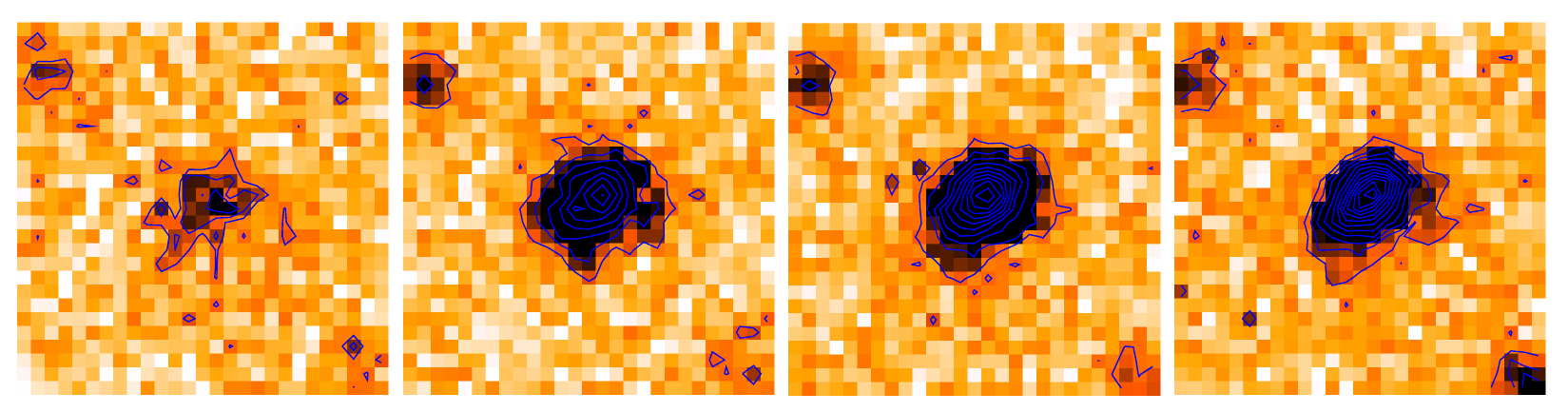}

\caption {Thumbnail images of VUDS 910298177 (seeing FWHM
$<0.75^{\prime\prime}$) $5^{\prime\prime}$ on a side, or $\sim$40
physical kpc at the background galaxy and DLA redshift.  Left to
right: the CFHTLS 5-year stacked $u^*$-, $g^\prime$-, $r^\prime$-, and
$i^\prime$-band images, respectively.  Linearly increasing flux
contours are overlaid in blue to help visualize the galaxy
morphology. The first (outermost) contour corresponds to the 1$\sigma$
sensitivity of the spectrum near the \lya\ absorption feature.}

\label{f2}
\end{center}
\end{figure}

Figure~\ref{f2} presents $5^{\prime\prime}\times5^{\prime\prime}$
CFHTLS image cutouts centered on \vvgal, with its properties listed in
Table 1.  The contours in Figure~\ref{f2} suggest a non-point-source,
asymmetric morphology that may be the result of either the extended
morphology of the background galaxy or a blending of two sources
corresponding to the background galaxy and emission from the DLA.  In
the latter case, the consistency of the image contours suggest that
the DLA may reside at a projected separation of
$\sim$1$^{\prime\prime}$ ($\sim$8 physical kpc) to the southeast of
the flux contour peaks.

\subsection{The $z\sim 2.4$ DLA in the spectrum of \vvgal}

We first estimate the H\textsc{i} content of EG1 by measuring its
equivalent width.  The feature is centered near $\lambda$ = 4124 \AA\
($z_{abs}$ = 2.391), and we choose a conservative wavelength window
2500 km s$^{-1}$ on either side (4088.1--4156.9 \AA) so as to avoid
contaminating absorption (specifically, the feature near $\lambda
\sim$ 4055 \AA).  These parameters result in a \lya\ rest equivalent
width $W_r \sim$ 17.0 \AA\ and a column density of log $N$(\hi) $\sim$
20.8 cm$^{-2}$.  We stress that this value underestimates the
H\textsc{i} content as the damping wings have been neglected.

Next, we use a Voigt profile model centered at $z_{abs}$ = 2.391 and
adopt the flux level of the composite spectrum as an estimate of the
continuum level.  We find that the feature is well modeled at log
N(\hi) $\sim$ 21.1 cm$^{-2}$.  Uncertainties resulting from the S/N,
low resolution, contaminating flux and absorption, and continuum level
allow for a range of model fits.  Thus, we conservatively adopt an
\hi\ content estimate of log $N$(\hi) = 21.1 $\pm0.4$ as illustrated
in Figure~\ref{f1}.

We stress that the models and our simulations of the feature {\it
require} that EG1 have this \hi\ column density covering $\gtrsim$90\%
of the background galaxy (i.e., a covering factor, $f_c$ $\gtrsim$
0.9).  If EG1 had an equivalent \hi\ column density but only covered a
small region of the background galaxy, the damping wings would not be
as pronounced and the core region would deviate significantly from
zero (i.e., much shallower than observed) because of the low
resolution of the VVDS-UltraDeep spectrum \citep{schaye01,heckman01}.

\section{The physical nature of EG1}\label{nature}

The size of \vvgal\ cannot be determined accurately using
seeing-limited ground-based data.  However, we can place meaningful
constraints on its size as follows.  The most physically compact
morphology to produce the star-forming luminosity of \vvgal\ (M $\sim$
$-$21) would be that of a highly dense $\sim$1 kpc super-star-forming
clump.  Such a compact background galaxy would probe continuous areas
of the DLAs $\gtrsim$100,000,000 larger than that probed by quasars
and gamma-ray-burst sightlines.  Galaxies at $z\sim$ 2--3 with
luminosities similar to, and fainter than, \vvgal\ have average
half-light radii of 1--3 kpc and a variety of morphologies -- from
single compact star-forming regions to extended diffuse emission with
multiple star-forming clumps -- and include systems with radii of
$\sim$10 kpc to the surface brightness limits of the deep
VVDS-UltraDeep spectroscopy
\citep{forster09,elmegreen05,law12a,law12b}.  If \vvgal\ is typical of
\zzz\ galaxies, the continuous area of EG1 probed would range from a
few to $\sim$100 kpc$^2$.  Previous research using the radio structure
of quasars, radio mapping, infrared adaptive optics observations, and
quasar close pair sightlines
\citep{briggs89,zwaan05,ellison07,monier09,cooke10,krogager12,jorgenson14}
suggests that DLAs span these extents.

Under the simple assumption that the DLA gas resides in a uniform
slab, the range of background galaxy sizes (1--100 kpc$^2$) and
1$\sigma$ column densities correspond to neutral gas masses of
$\sim$10$^{6}$--10$^{9}$ $M_\odot$ using ${(\hi(cm^{-2})\times
A(cm^2)\times m_{proton}(g))/1.98\times 10^{33}g/M_\odot}$.

The mass estimate only considers the area probed by the background
galaxy flux and, because gas densities in DLA clouds should diminish
toward their outer radii, a significant fraction of the gas may extend
over larger areas.  The mass estimate is consistent with DLA gas
associated with \zzz\ DLA dark matter halos from observation
\citep{cooke06} and theory \citep[e.g.,][]{fumagalli11,rahmati14}.

Such large reservoirs of self-shielded neutral gas are protogalaxies,
or extended components of established galaxies, and are expected to
produce significant star formation as the system evolves.  Inspection
of the images reveals a subtle elongation in the shape of \vvgal\ and,
interestingly, faint flux may be present in the 2D spectrum offset
from the \lya\ absorption (below the spectrum as shown in
Figure~\ref{f1}) and spatially consistent with the elongation.  If
real, these features may result from emission by EG1 and imply that it
is a faint galaxy.

To test this hypothesis, we conservatively assume that the background
galaxy produces zero Lyman continuum flux (which, here, falls in the
$u^*$-band) following the Lyman break expectations
\citep{steidel93,steidel03}.  We note that some \zzz\ galaxies are
theorized, and observed, to exhibit Lyman continuum flux of a few
percent as compared to their flux near 1500\AA\
\citep{steidel01,cooke14}.  A photometric and spectroscopic comparison
of \vvgal\ finds that the slit spectroscopy acquired $\sim$52\% of the
total galaxy flux in the $g^\prime$$r^\prime$$i^\prime$ filters.  We
apply this correction to the $u^*$-band and conservatively assign all
excess $u^*$-band flux of the background galaxy to the DLA.  This
approach yields a DLA brightness of m(1000 \AA) $\sim$ 27.1,
corresponding to roughly m(1500 \AA) = 26.3, a star formation rate of
$\sim$2 $M_\odot$ yr$^{-1}$, and a halo mass $M_{DM}\sim$10$^{11}$
$M_\odot$ \citep{berrier12}.  These upper limits are consistent with
the offset flux observed in the images, the values found in DLA
emission searches in quasar sightlines
\citep{moller02,fynbo11,peroux12,krogager12,jorgenson14,fumagalli14},
and the number density of $m_{i\prime}\sim$ 27 \zzz\ galaxies
\citep{schaye01}.  The above test demonstrates that an interpretation
of EG1 as a high-redshift galaxy is plausible (see also \S\ref{LAA}).

\section{The occurrence rate of DLAs in galaxy spectra}

As a first step to estimate the occurrence rate of DLAs in galaxy
spectra, we assess the frequency of absorbers having similar \hi\
column densities as EG1 in quasar spectra.  The \vvgal\ spectrum
provides a redshift path $z=$ 1.96--2.81 to search for DLA \lya\
absorption, corresponding to a cosmological pathlength of ${\Delta
X=2.768}$.  The DLA incidence frequency is $\ell(X)$, where the number
of DLAs in pathlength $\Delta X$ is given by $\ell(X)\Delta X$.  If we
adopt the $z\sim$ 2.5 H\textsc{i} column density distribution function
of \citet{noterdaeme12}, we find $\ell(X)=$ 0.025 for DLAs in the
allowed N(\hi) range of EG1 (log $N$(\hi) = 21.1 $\pm$0.4 cm$^{-2}$).
Thus, we would have to observe $\sim$14 quasars probing redshift range
${z= 1.96-2.81}$ to detect one DLA with the same H\textsc{i} content.
If we restrict ourselves to the range log $N$(H\textsc{i}) 21.1
$\pm$0.05 cm$^{-2}$, the number of quasars needed increases to
$\sim$150, owing to the steep shape of the H\textsc{i} column density
distribution function.  Thus, while uncommon, absorbers like EG1 are
not exceedingly rare when observed in quasar spectra.  Current
simulations indicate that absorbers with sizes and column densities
similar to EG1 comprise $\sim$1\% of the DLA population
\citep{fumagalli11,rahmati14}.

While the full analysis for our program sample will appear in later
papers, we can make initial estimates of the incidence frequency of
DLAs in galaxy sightlines.  From the 54 secure VVDS-UltraDeep LCG
spectra that we assessed, we find two additional DLA candidate systems
($W_r$(\lya) $>5$ \AA), providing a coarse $\sim$6\% occurrence rate
estimate for systems with log $N$(H\textsc{i}) $\gtrsim$ 20.3
cm$^{-2}$ and $\sim$2\% for systems with log $N$(H\textsc{i})
$\gtrsim$ 21.1 cm$^{-2}$.  Depending on significance level, we find
$\sim$6 DLA candidate systems in the lower S/N Keck programs LBG
spectra.  Under the assumption that these $\sim$6 systems are DLAs,
their occurrence rates are $\sim$3--5$\times$ lower than the
VVDS-UltraDeep LCGs.

As described in \S\ref{nature}, excess Lyman continuum flux in LCGs
with DLAs could result from DLA galaxy emission.  If the difference in
the estimated LCG and LBG occurrence rates persists with larger
statistical samples, the results would support the scenario in which
DLAs often produce detectable flux and, thus, are associated with
star-forming galaxies
\citep{moller02,cooke06,fynbo11,peroux12,krogager12,jorgenson14}.
DLAs in galaxy sightlines provide opportunities to measure their host
galaxy properties without the glare typically present from bright
quasars.

\section{Galaxies with dominant \lya\ in absorption and the nature of
the neutral gas}\label{LAA}

Previous work analysing $r\lesssim$ 25.5 LBG spectra
\citep[e.g.,][]{shapley03,cooke06} find that roughly 25\% exhibit
dominant \lya\ in absorption, $\sim$25\% show dominant \lya\ in
emission, and the remainder exhibit a combination of \lya\ emission
and absorption.  Many LBGs with dominant \lya\ absorption show
evidence for damped absorption \citep{pettini00,shapley03}.  Finally,
a population of faint, high-redshift \lya\ emitting galaxies (LAEs)
have been identified via the detection of \lya\ emission in narrowband
imaging surveys \citep[e.g.,][]{hu98} and appear to be similar to the
$\sim$25\% of LBGs that show \lya\ in emission but have bluer UV
continuum slopes and lower star formation rates
\citep{gawiser06,cooke09,cooke13,garel15}.

Spectra of faint ($r\gtrsim$ 25.5) LBGs with an S/N of more than a few
are difficult to obtain with 8 m class telescope in reasonable
integration times.  The spectra acquired to date show a trend for a
larger fraction of galaxies showing \lya\ in emission.  This trend may
be an observational bias because very faint galaxy spectra are easier
to identify when \lya\ emission is present.  As a result, the
identification of r $\gtrsim$ 25 galaxies with dominant \lya\ in
absorption (LAAs) and their properties remain largely unknown.  DLAs
randomly sample all galaxies with sufficient neutral gas, including
very faint and low-mass galaxies.  Here, we examine the likelihood
that EG1 is an LAA.

\begin{figure}
\begin{center}
\includegraphics[width=0.55\textwidth,angle=90]{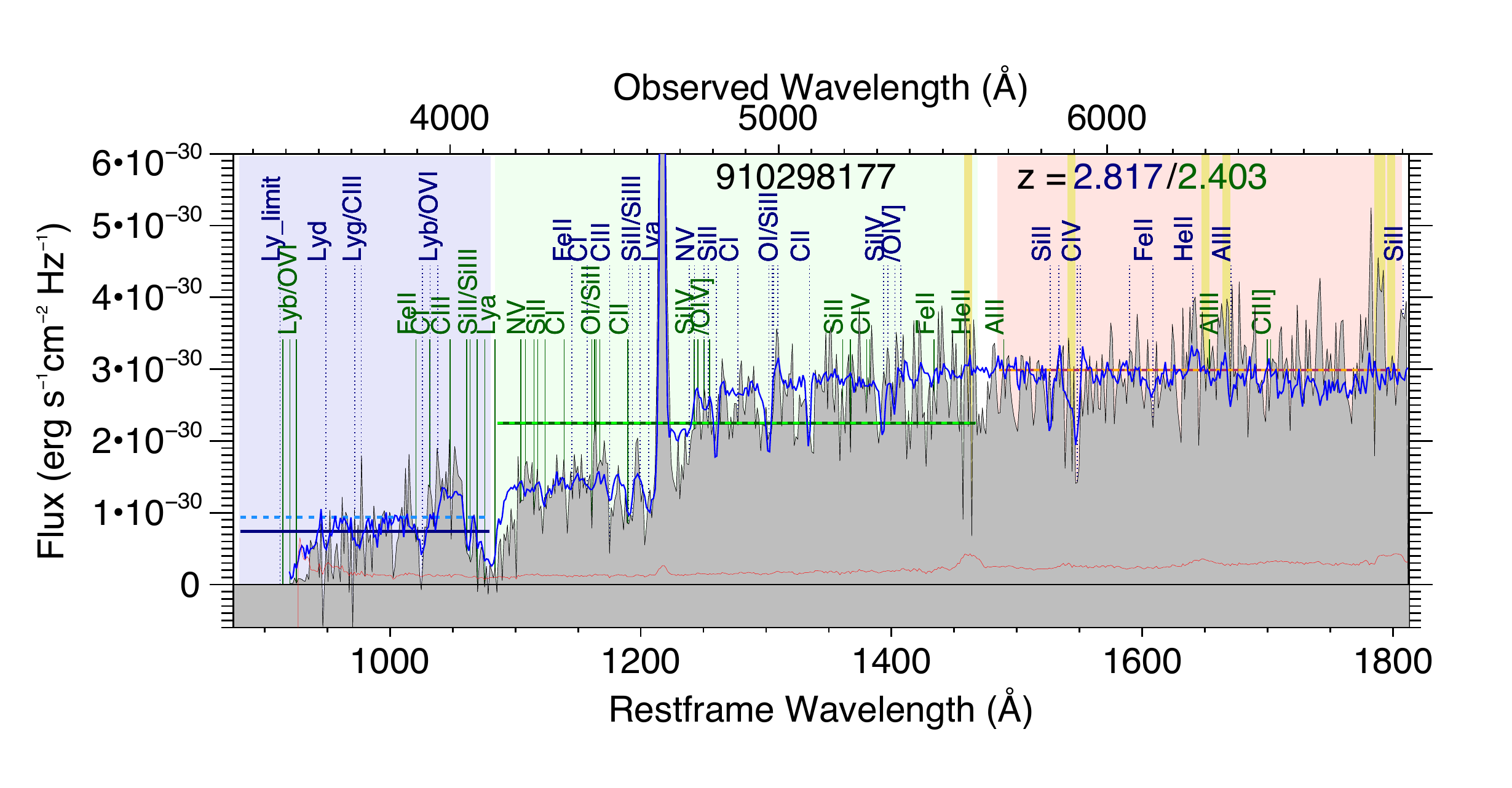}

\caption {One-dimensional spectrum of EG1 and background galaxy
plotted similarly to Figure~\ref{f1}.  The composite spectrum is
overlaid (blue) is composed of $\sim$200 \zzz\ LBGs with dominant
\lya\ in emission (eLBG spectrum) at the redshift of the background
galaxy ($z=$ 2.817) and $\sim$200 LBGs with dominant \lya\ in
absorption (aLBG spectrum) at the redshift of EG1 ($z=$ 2.403).  The
eLBG spectrum is a good fit to the background galaxy, as shown in
Figure~\ref{f1}, and the aLBG spectrum is added to model EG1.
Prominent atomic transitions for the background galaxy (blue labels,
upper row) and EG1 (green labels, lower row) are indicated.  Solid
horizontal lines represent the photometric flux levels for the $u^*$,
$g^\prime$, and $r^\prime$ filter bandpasses (blue, green, and red,
respectively) using the $r^\prime$-band correction factor {\it (see
the text)}.  The dashed horizontal lines denote the mean spectrum flux
over that bandpass. }

\label{f3}
\end{center} 
\end{figure}

To model the \vvgal\ spectrum as two galaxies, we first use a
composite spectrum constructed from $\sim$200 $r <$ 25.5 LBG spectra
with dominant \lya\ in emission to model the background galaxy
(hereafter eLBG spectrum).  To model EG1, we use a composite spectrum
constructed from $\sim$200 $r <$ 25.5 LBGs with dominant \lya\ in
absorption (hereafter aLBG spectrum).  We place the eLBG spectrum at
the redshift of the background galaxy and scale it to the 1500 \AA\
flux of the background galaxy minus the upper limit flux of EG1
($r^\prime$ = 26.3).  We then overlay and scale the aLBG spectrum to
the redshift and 1500 \AA\ flux of EG1.  Finally, we note that EG1 is
in the foreground and its damped \lya\ absorption feature will absorb
the background galaxy flux (i.e., not be convolved with it).  We scale
the template to reflect a $\sim$90\% absorption to match the lower
limit of the data.  Figure~\ref{f3} shows the resulting composite
spectrum (eLBG+aLBG) overlaid onto the data.

The \lya\ feature of the aLBG spectrum is in very good agreement with
the \lya\ feature of EG1 in profile, depth, and width (i.e., EG1 is
similar to an LAA) when considered in the proper context.  The \lya\
absorption in the composite aLBG spectrum results largely from
absorption foreground to star-forming regions in the LBGs that compose
it.  In contrast, all the gas in EG1 absorbs the background galaxy,
including advancing and receding gas as a result of outflows.  The red
portion of the aLBG spectrum \lya\ feature does not include absorption
from receding gas nor the circumgalactic material on the far side of
EG1 and includes \lya\ flux from resonantly scattered photons of the
LBGs with higher star formation rates that compose it.

Interestingly, \vvgal\ exhibits a strong absorption feature blueward
of EG1 \lya\ similar to the Si \textsc{ii} 1193 \AA, Si \textsc{iii}
1200, 1207 \AA\ feature in the aLBG spectrum, and there is low S/N
evidence for Si \textsc{ii} 1260, 1304, 1527 \AA\ and Si \textsc{iv}
1393, 1403 \AA\ absorption by EG1.  If real, this behavior implies a
large Si covering fraction for EG1 and similar properties to $r
\lesssim$ 25.5 LAAs.

Finally, we test the assumption that DLAs are galaxies like LBGs using
their number density.  Previous work shows that the number density of
DLAs and \zzz\ $r \lesssim$ 27 LBGs (i.e., n = 0.016 {\it
h}$^3$Mpc$^{-3}$; $M_{DM}\gtrsim$ 10$^{11}$ $M_\odot$) are consistent
if DLAs have average radii of 19 kpc \citep{schaye01}.  However, most
DLAs are likely fainter than $r \lesssim$ 27, as our sample and
unbiased searches indicate \citep{fumagalli14}.

If, instead, we assume DLAs have average radii of $\sim$5.7 kpc (i.e.,
average areas of 100 kpc$^2$), they would sample $r^\prime$ $\lesssim$
30 LBGs (M $\sim$ -15.5; $M_{DM}\gtrsim$ 10$^{10.7}$ $M_\odot$) with
an extrapolation of the LBG luminosity function
\citep[e.g.,][]{reddy09}.  This speculation is plausible, given the
estimated areas of DLAs in our sample and because log $N$(H\textsc{i})
$>$ 20.3 gas likely extends to areas larger than that probed by the
background galaxies.

Finally, we note that UV-luminous LBGs do not comprise all galaxies at
high redshift, and there is evidence that they may constitute as
little as $\sim$15--20\% \citep{berrier12,spitler14}.  Because DLAs
randomly probe galaxies independent of their UV luminosities, average
radii of $\sim$5.7 kpc would probe $r \lesssim$ 28 galaxies at \zzz\
in such a scenario, allowing much smaller average DLA radii when
probing average galaxies by number.

\section{The Future}

The discovery of high-redshift DLAs in sightlines to galaxies provides
the first means to uncover their true sizes.  Our approach of
searching LBG and LCG sightlines exploits commonly occurring galaxies,
as opposed to rare quasars and gamma ray bursts, while probing
continuous areas more than 10 million times larger.  Future analysis
of our full DLA sample will include background galaxy sizes determined
from Cycle 23 {\it Hubble Space Telescope} imaging (GO 14160; PI:
O'Meara).

The general approach pushes the capabilities of 8 m class telescope
facilities to their limits.  However, searches for DLAs in the
sightlines to background galaxies will likely become the dominant
method of study in the upcoming era of 30 m class telescopes.  Nominal
first light instruments enable ${R=5000}$ spectroscopy of m $<$ 24.5
galaxies at S/N $>$ 35 in 4.5 hr (e.g., WFOS on the Thirty Meter
Telescope).  The increase in both resolution and S/N would enable more
accurate $N$(\hi) and metallicity measurements.  In addition,
lower-resolution spectroscopy ($R \sim$ 1000), enables $>$5$\sigma$
DLA detection in m $<$ 26.5 galaxy spectra in 1 hr.  Probing to
fainter galaxies dramatically increases the number density, enabling a
tomographic neutral gas mass reconstruction of the early Universe,
while helping to complete our understanding of galaxy formation and
evolution.\\


JC thanks Olivier Le F{\`e}vre and Michael Murphy for helpful
discussions.  JC acknowledges support from Australian Research Council
grant FF130101219.  JMO thanks the Swinburne Visiting Scientist scheme
that helped make this work possible.  This research uses data from the
VIMOS VLT Deep Survey, obtained from the VVDS database operated by
Cesam, Laboratoire d'Astrophysique de Marseille, France.  In addition,
some of the data presented here are based on observations obtained
with MegaPrime/MegaCam, a joint project of CFHT and CEA/IRFU, at the
Canada-France-Hawaii Telescope (CFHT) which is operated by the
National Research Council (NRC) of Canada, the Institut National des
Science de l'Univers of the Centre National de la Recherche
Scientifique (CNRS) of France, and the University of Hawaii.  This
work is based in part on data products produced at Terapix available
at the Canadian Astronomy Data Centre as part of the
Canada-France-Hawaii Telescope Legacy Survey, a collaborative project
of NRC and CNRS.  Finally, some of the data presented herein were
obtained at the W.M. Keck Observatory, which is operated as a
scientific partnership among the California Institute of Technology,
the University of California and the National Aeronautics and Space
Administration. The Observatory was made possible by the generous
financial support of the W.M. Keck Foundation.  The authors wish to
recognize and acknowledge the very significant cultural role and
reverence that the summit of Mauna Kea has always had within the
indigenous Hawaiian community.  We are most fortunate to have the
opportunity to conduct observations from this mountain.

\bibliographystyle{apj}

\newpage
\begin{table}\label{table1}
\begin{tabular}{lcccccc}
\hline VVDS-UltraDeep Name & R.A. (J2000) & Decl. (J2000) & $m_{u^*}$ 
& $m_{g^\prime}$ & $m_{r^\prime}$ & $m_{i^\prime}$\\
& h m s & d m s & (AB) & (AB) & (AB) & (AB)\\
\hline
VVDS 910298177 & 02 26 54.699 & -04 18 45.77 & $25.699\pm0.091$ &
$24.751\pm0.037$ & $24.543\pm0.035$ & $24.535\pm0.040$ \\
\hline
\hline
\end{tabular}

\caption{Properties of \vvgal}
\label{sal}
\end{table}

\newpage

\end{document}